Title: Watson-Crick Quantum Finite Automata

Authors: Kingshuk Chatterjee[1], Kumar Sankar Ray[2]

Affiliations: [1,2]Electronics and Communication Sciences Unit, Indian Statistical Institute, Kolkata-108.

# Watson-Crick Quantum Finite Automata

*Kingshuk Chatterjee[1], Kumar Sankar Ray[2]*

[1,2]*Electronics and Communication Sciences Unit, Indian Statistical Institute, Kolkata-108.*

[1]kingshukchaterjee@gmail.com  [2]ksray@isical.ac.in

*Abstract: one-way quantum finite automata are reversible in nature, which greatly reduces its accepting property. In fact, the set of languages accepted by one-way quantum finite automata is a proper subset of regular languages. In this paper, we replace the tape head of one-way quantum finite automata with DNA double strand and name the model Watson-Crick quantum finite automata. The non-injective complementarity relation of Watson-Crick automata introduces non-determinism in the quantum model. We show that this introduction of non-determinism increases the computational power of one-way Quantum finite automata significantly. We establish that Watson-Crick quantum finite automata can accept all regular languages and that it also accepts some languages which are not accepted by any multi-head deterministic finite automata. Exploiting the superposition property of quantum finite automata we show that Watson-Crick quantum finite automata accept the language L={ww |w∈ $\{a, b\}^*$}.*

## I. INTRODUCTION

Kondacs et.al.[1] and Moore et.al.[2] introduced two different models of quantum finite automata. Kondacs et.al. gave more emphasis to two-way quantum automata because they are more powerful. Ambainis et. al. discussed in details one-way model of Quantum finite automata(1QFA)[3]. One-way quantum finite automata are the simplest model of quantum automata. As already stated, two different models of one-way quantum finite automata are in existence, viz, the measure many one-way quantum finite automata [1] and the measure once one-way quantum finite automata [2]. Ambainis et. al. showed that measure many one-way quantum finite automata can accept all languages that can be accepted by measure once one-way quantum finite automata. Hence, in this paper, we consider the measure many quantum finite automata described by Kondacs et. al. [1]. Whenever we mention one-way quantum finite automata we mean the model described by Kondacs et. al. It has been shown by Kondacs et. al. [1] that one-way Quantum finite automata are a proper subset of regular languages.

Watson-Crick automata [4] are finite automata having two independent heads working on double strands where the characters on the corresponding positions of the two strands are connected by a complementarity relation similar to the Watson-Crick complementarity relation. The movement of the heads although independent of each other is controlled by a single state. Paun et.al.[5] explored several variants of non-deterministic Watson-Crick automata (AWK). Czeizler et.al.[6] discussed the computational power of deterministic Watson-Crick automata. Equivalence of subclasses of two-way Watson-Crick automata is discussed in [7]. A survey of Watson-Crick automata can be found in [8]. Research work regarding state complexity of Watson-Crick automata is reported in [9] and [10].

In this paper, we introduce a new model of one-way quantum finite automata, namely Watson-Crick quantum finite automata (1WKQFA) which are one-way quantum finite automata where the input tape is replaced by double strands with two independent heads. Quantum finite automata are inherently reversible in nature and it is this absence of non-determinism in the model which greatly reduces their computational power. The double strand and the non-injective complementarity relation in the Watson-Crick quantum finite automata enable us to introduce non-determinism in the quantum model. We primarily shift the non-determinism from the automaton to the input. Thus, the automaton still remains reversible and retains its quantum properties. We show that Watson-Crick quantum finite automata in spite of being reversible accept all regular language. We further show that the above mentioned model is more powerful than one-way reversible 2-head finite automata. In fact, Watson-Crick quantum finite automata accept languages which are not accepted by any multi-head deterministic finite automata. Moreover by exploiting the superposition property, we show that Watson-Crick quantum finite automata accept the language L={ww |w∈ $\{a, b\}^*$}.

## II. BASIC TERMINOLOGY

The symbol V denotes a finite alphabet. The set of all finite words over V is denoted by $V^*$, which includes the empty word $\lambda$. The symbol $V^+=V^*-\{\lambda\}$ denotes the set of all non-empty words over the alphabet V. For w ∈ $V^*$, the length of w is denoted by |w|. Let u∈ $V^*$ and v ∈ $V^*$ be two words and if there is some word x ∈ $V^*$, such that v=ux, then u is a prefix of v, denoted by u ≤ v. Two words, u and v are prefix comparable denoted by u~$_p$v if u is a prefix of v or vice versa.

### Watson-Crick automata

A Watson-Crick automaton is a 6-tuple of the form M=(V,$\rho$,Q,$q_0$,F,$\delta$) where V is an alphabet set, set of states is denoted by Q, $\rho \subseteq V \times V$ is the complementarity relation similar to Watson-Crick complementarity relation, $q_0$ is the initial state and F⊆Q is the set of final states. The function $\delta$ contains a finite number of transition rules of the form q$\binom{w_1}{w_2}$→q', which denotes that the machine in state q parses $w_1$ in upper strand and $w_2$ in lower strand and goes to state q' where $w_1, w_2 \in V^*$. The symbol $\begin{bmatrix} w_1 \\ w_2 \end{bmatrix}$ is different from $\begin{pmatrix} w_1 \\ w_2 \end{pmatrix}$. While $\begin{pmatrix} w_1 \\ w_2 \end{pmatrix}$ is just a pair of strings written in that form instead of $(w_1,w_2)$, the symbol $\begin{bmatrix} w_1 \\ w_2 \end{bmatrix}$ denotes that the two strands are of same length i.e. $|w_1|=|w_2|$ and the corresponding symbols in two strands are complementarity in the sense given by the relation ρ. The symbol $\begin{bmatrix} V \\ V \end{bmatrix}_\rho = \{\begin{bmatrix} a \\ b \end{bmatrix} \mid a, b \in V, (a, b) \in \rho\}$ and $WK_\rho(V) = \begin{bmatrix} V \\ V \end{bmatrix}_\rho^*$ denotes the Watson-Crick domain associated with V and ρ.

A transition in a Watson-Crick finite automaton can be defined as follows:

For $\begin{pmatrix} x_1 \\ x_2 \end{pmatrix}, \begin{pmatrix} u_1 \\ u_2 \end{pmatrix}, \begin{pmatrix} w_1 \\ w_2 \end{pmatrix} \in \begin{pmatrix} V^* \\ V^* \end{pmatrix}$ such that $\begin{bmatrix} x_1 u_1 w_1 \\ x_2 u_2 w_2 \end{bmatrix} \in WK_\rho(V)$ and q, q' $\in Q$, $\begin{pmatrix} x_1 \\ x_2 \end{pmatrix} q \begin{pmatrix} u_1 \\ u_2 \end{pmatrix} \begin{pmatrix} w_1 \\ w_2 \end{pmatrix} \Rightarrow \begin{pmatrix} x_1 \\ x_2 \end{pmatrix} \begin{pmatrix} u_1 \\ u_2 \end{pmatrix} q' \begin{pmatrix} w_1 \\ w_2 \end{pmatrix}$ iff there is transition rule $q \begin{pmatrix} u_1 \\ u_2 \end{pmatrix} \to q'$ in δ and $\stackrel{*}{\Rightarrow}$ denotes the transitive and reflexive closure of ⇒. The language accepted by a Watson-Crick automaton M is $L(M) = \{w_1 \in V^* \mid q_0 \begin{bmatrix} w_1 \\ w_2 \end{bmatrix} \stackrel{*}{\Rightarrow} q \begin{bmatrix} \lambda \\ \lambda \end{bmatrix}$, with $q \in F, w_2 \in V^*, \begin{bmatrix} w_1 \\ w_2 \end{bmatrix} \in WK_\rho(V)\}$.

**Deterministic Watson-Crick Automata**

The notion of determinism in Watson-Crick automata and a discussion on its complexity were first considered in [6]. In [6] different notions of determinism were suggested as follows:

1) weakly deterministic Watson-Crick automata(WDWK): Watson-Crick automaton is weakly deterministic if in every configuration that can occur in some computation of the automaton, there is a unique possibility to continue the computation, i.e. at every step of the automaton there is at most one way to carry on the computation.

2) deterministic Watson-Crick automata(DWK): deterministic Watson-Crick automaton is Watson-Crick automaton for which if there are two transition rules of the form $q\begin{pmatrix} u \\ v \end{pmatrix} \to q'$ and $q\begin{pmatrix} u' \\ v' \end{pmatrix} \to q''$ then $u \not\sim_p u'$ or $v \not\sim_p v'$.

3) strongly deterministic Watson-Crick automata(SDWK): strongly deterministic Watson-Crick automaton is a deterministic Watson-Crick automaton where the Watson-Crick complementarity relation is injective.

**One-way Quantum Finite automata**

One-way quantum finite automaton is a six tuple $M=(Q, V, \delta, q_0, Q_{acc}, Q_{rej})$ where Q is a finite set of states, V is the input alphabet, δ is the transition function, $q_0 \in Q$ is a starting state and $Q_{acc} \subset Q$, and $Q_{rej} \subset Q$ are sets of accepting and rejecting states. The states in $Q_{acc}$ and $Q_{rej}$ are called halting states and the states in $Q_{non}=Q-(Q_{acc} \cup Q_{rej})$ are called the non-halting states. '#' and '$' are symbols that do not belong to V. We use '#' and '$' as the left and right endmarkers respectively. The working alphabet of M is $\Gamma=V \cup \{\#,\$\}$.

A superposition of M is any element of $l_2(Q)$. For $q \in Q$, $|q\rangle$ denotes the unit vector with value 1 at q and 0 elsewhere. All elements of $l_2(Q)$ can be expressed as a linear combination of vectors $|q\rangle$. We will use ψ to denote $l_2(Q)$.

The transition function δ maps $Q \times \Gamma \times Q$ to C where C denotes the set of complex numbers. The value $\delta(q_1,a,q_2)$ is the amplitude of $|q_2\rangle$ in the superposition of states to which M goes from $|q_1\rangle$ after reading 'a'. For $a \in \Gamma$, $U_a$ is a linear transformation on $l_2(Q)$ defined by $U_a(|q_1\rangle) = \sum_{q_2 \in Q} \delta(q_1, a, q_2)|q_2\rangle$.

We require all $U_a$ to be unitary.

The computation of a one-way quantum finite automaton starts in the superposition $|q_0\rangle$. Then transformations corresponding to left endmarker '#', the letters of the input word w and the right endmarker '$' are applied.

The transformation corresponding to $a \in \Gamma$ consists of two steps.

1) First, $U_a$ is applied. The new superposition ψ' is $U_a(\psi)$ where ψ is the superposition before this step.

2) Then, ψ' is observed with respect to the observable $E_{acc} \oplus E_{rej} \oplus E_{non}$ where $E_{acc}=span\{|q\rangle:q \in Q_{acc}\}$, $E_{rej}=span\{|q\rangle:q \in Q_{rej}\}$, $E_{non}=span\{|q\rangle:q \in Q_{non}\}$. This observation gives $x \in E_i$ with probability equal to the amplitude of the projection of ψ'. After that the superposition collapses to the projection.

If we get ψ'∈ $E_{acc}$, the input is accepted. If ψ'∈ $E_{rej}$, the input is rejected. If ψ'∈ $E_{non}$, the next transformation is applied.

We regard these two transformations as reading a letter 'a'.

The above stated definition of 1QFA is from [1].

For further clarity of the above mentioned definition and notations of 1QFA we may consider the definition in Section 6 and Example in Section 3 given in [1].

### III. WATSON-CRICK QUANTUM FINITE AUTOMATA

Watson-Crick quantum finite automaton is a seven tuple $M=(Q, V, \delta, q_0, Q_{acc}, Q_{rej}, \rho)$ where Q is a finite set of states, V is the input alphabet, δ is the transition function, $q_0 \in Q$ is a starting state and $Q_{acc} \subset Q$, and $Q_{rej} \subset Q$ are sets of accepting and rejecting states. The complementarity relation ρ is similar to Watson-Crick complementarity relation. The states in $Q_{acc}$ and $Q_{rej}$ are called halting states and the states in $Q_{non}=Q-(Q_{acc} \cup Q_{rej})$ are called the non-halting states. The symbols '#' and '$' do not belong to V. We use '#' and '$' as the left and right endmarkers respectively. The working alphabet of M is $\Gamma=V \cup \{\#,\$\}$. The input tape is a double stranded input tape with two heads each on one of the strands where the letters in the corresponding positions on the input tape are according to the complementarity relation ρ. The word on the upper strand is accepted or rejected by the automaton.

A superposition of M is any element of $l_2(Q)$. For $q \in Q$, $|q\rangle$ denotes the unit vector with value 1 at q and 0 elsewhere. All elements of $l_2(Q)$ can be expressed as a linear combination of vectors $|q\rangle$. We will use ψ to denote $l_2(Q)$.

The transition function δ maps $Q \times \Gamma^2 \times Q \times \{0,1\}^2$ to C where C denotes the set of complex numbers. The value $\delta(q_1,a,b,q_2,d_1,d_2)$ is the amplitude of $|q_2\rangle$ in the superposition of states to which M goes from $|q_1\rangle$ after reading 'a' in the upper strand and 'b' in the lower strand and moving the upper head according to $d_1$ and lower head according to $d_2$ where zero denotes head stays in its position and one denotes head has moved to the right. For $a, b \in \Gamma$, $U_{a,b}$ is a linear transformation on $l_2(Q)$ defined by $U_{a,b}(|q_1\rangle) = \sum_{q_2 \in Q} \delta(q_1, a, b, q_2, d_1, d_2)|q_2\rangle$. We require all $U_{a,b}$ to be unitary. The check for well-formedness can be done in a similar manner as in [1] in the following way:

Consider the Hilbert space $l_2(Q)$, where Q is the set of internal states of the automaton M. A linear operator $U_{\sigma,\tau}: l_2(Q) \to l_2(Q)$ for each σ,τ pair and a function $D: Q \to \{0,1\}^2$ exist. The transition function δ is defined as $\delta(q, \sigma, \tau, q', d_1, d_2) = \begin{cases} \langle q'|U_{\sigma,\tau}|q\rangle & D(q') = (d_1, d_2) \\ 0 & D(q') \neq (d_1, d_2) \end{cases}$

where $\langle q'|U_{\sigma,\tau}|q\rangle$ denotes the coefficient of $|q'\rangle$ in $U_{\sigma,\tau}|q\rangle$. M is well-formed if and only if $\sum_{q'} \overline{\langle q'|U_{\sigma,\tau}|q_1\rangle} \langle q'|U_{\sigma,\tau}|q_2\rangle = \begin{cases} 1 & q_1 = q_2 \\ 0 & q_1 \neq q_2 \end{cases}$ for each σ,τ pair. The condition mentioned is similar to the condition for reversibility in [11].

The input word w is of the form $\begin{bmatrix} w_1 \\ w_2 \end{bmatrix} \in WK_\rho(V)$, where the automaton accepts or rejects $w_1$ with some probability. Both strands begin with # and ends with $. The string $\#w_1\$$ is placed in the upper strand and $\#w_2\$$ in the lower strand.

Note that many values of $U_{\sigma,\tau}|q\rangle$ define transitions which we do not encounter during a computation of w for a particular M. We define those values arbitrarily in such a way that $U_{\sigma,\tau}$ is unitary. In general we specify only those values that matter for all other values the automaton M goes to some state q where $q \in Q$, the other values are so assigned that the resulting operator is unitary. So for a state q if no value is mentioned for a pair σ,τ where σ,τ $\in \Gamma$, as Γ is finite therefore number of such σ,τ pairs are also finite. $U_{\sigma,\tau}|q\rangle = |q_{rejq}\rangle$ where defining unmentioned transitions in this way ensures well-formed transitions. Moreover for a given automaton M if such a transition is employed it always rejects. This enables us to define automaton without mentioning all the σ,τ pairs. We assume these $q_{rejq}$'s belongs to the set $Q_{rej}$ and $D(q_{rejq}) = (0,0)$. As these transitions are included in automaton by default when mentioning the set $Q_{rej}$ and Q we do not explicitly mention these $q_{rejq}$'s.

A Watson-Crick quantum finite automaton(WKQFA) is called a ***strongly Watson-Crick quantum finite automaton*** (SWKQFA) if the complementarity relation is injective in that particular automaton.

**Example 1**: M=(Q, V, δ, $q_0$, $Q_{acc}$, $Q_{rej}$,ρ) is a strongly deterministic Watson-Crick quantum finite automaton that accepts the context sensitive language $a^n b^n c^n$ n≥1 where Q={$q_0,q_1,q_2,q_3,q_{acc}$}, $Q_{acc}$={$q_{acc}$}, $Q_{rej}$={}, V={a,b,c}, ρ is the injective complementarity relation. We define the linear operator in M as follows.

$U_{\#,\#}|q_0\rangle=|q_0\rangle$, $U_{\#,a}|q_0\rangle=|q_0\rangle$, $U_{\#,b}|q_0\rangle=|q_1\rangle$, $U_{a,b}|q_1\rangle=|q_1\rangle$, $U_{a,c}|q_1\rangle=|q_2\rangle$, $U_{b,c}|q_2\rangle=|q_2\rangle$, $U_{b,\$}|q_2\rangle=|q_3\rangle$, $U_{c,\$}|q_3\rangle=|q_3\rangle$, $U_{\$,\$}|q_3\rangle=|q_{acc}\rangle$, $D(q_0)=(0,1)$, $D(q_1)=(1,1)$, $D(q_2)=(1,1)$, $D(q_3)=(1,0)$, $D(q_{acc})=(0,0)$.

By inspection we see that $U_{\sigma,\tau}$ is well-formed. The automaton checks the number of a's in the upper strand with the number of b's in the lower strand again repeats the procedure for number of b's and c's. The above automaton accepts a string in the language with probability 1 and also rejects a string not in the language with probability 1.

### IV. COMPUTATIONAL COMPLEXITY OF WATSON-CRICK QUANTUM FINITE AUTOMATA

**Theorem 1:** Strongly deterministic Watson-Crick quantum finite automata can accept all unary regular languages.

Proof: Kutrib et. al. stated that one-way multi-head reversible finite automata with two heads (1RMFA(2)) accept all unary regular languages. From the definition of strongly Watson-Crick quantum finite automata it is evident that if no superposition of states are involved in the strongly Watson-Crick quantum finite automaton, then the automaton behaves like a one-way reversible finite automata with two heads. As a result strongly Watson-Crick quantum finite automata also accept all unary languages.

**Theorem 2:** For every deterministic finite automaton which accepts a language L, we can find a Watson-Crick quantum finite automaton which accepts the same language (L).

**Proof:** The proof of the above Theorem is in two parts. In the first part given a deterministic finite automaton M which accepts a language L, we give a construction to obtain a Watson-Crick quantum finite automaton M' from M and in the second part we show that M' accepts the same language as M.

**First Part**: Given a deterministic finite automaton M= (Q, V, $q_0$, F, δ). We construct a Watson-Crick automaton Quantum finite automaton M'=(Q', V', δ', $q_0$', $Q_{acc}$, $Q_{rej}$, ρ) from M in the following manner:

For every x∈V, we do the following steps:

1) We form a list of all the transitions in M involving x, where these transitions involving x are arranged in any particular order and each transition is assigned a number of the form $x_i$ based on its position in the list. i.e. a transition is assigned a number $x_i$, if the transition is the $i^{th}$ transition in the list for x∈V.

2) Let us suppose there are n transitions in the list, then we introduce the symbols $x,x_1,...,x_n$ in V' and the relations $(x,x_1),(x,x_2),...,(x,x_n)$ in ρ.

3) For a transition δ(q,x)=q' having number $x_i$ associated with it, we introduce the transition $U_{x,x_i}|q\rangle=|q'\rangle$ in δ'.

Step 3 is repeated for every transition in the list.

Moreover the following transitions are also added to δ'.
1) $U_{\#,\#}|q_0'\rangle=|q_0\rangle$
2) $U_{\$,\$}|q\rangle=|q_{acc}\rangle$ for all q∈F

The set of states of M' i.e. Q'=Q∪{$q_0$', $q_{acc}$}.
The set of accepting states of M' i.e. $Q_{acc}$={$q_{acc}$}.
The set of rejecting states of M' i.e. $Q_{rej}$={}.
The start state of M' is $q_0$'.
Moreover, D(q)=(1,1) for all q∈Q'-$Q_{acc}$-$Q_{rej}$ and D($q_{acc}$)=(0,0).

As already mentioned in the definition of Watson Crick quantum finite automata all transitions which are not defined in δ' goes to the some rejecting state.

**Second Part:** In this part we show that M' constructed from M accepts the same language as M. Suppose M accepts w. The complementarity relation ρ of M' is so designed that the complementarity string w' of w guesses the transitions that M takes to accept w. Each position of w' guesses the transition that M takes on reading that particular position in w. Based on the sequence in w', M' simulates the transition sequence of M. As M accepts w, there is a sequence of transitions that takes M to its final state after consuming w. Thus, one of the many complementarity strings of w will rightly guess that particular sequence of transitions that enables M to accept w and for that particular complementarity string as M' simulates M based on the complementarity string of w; M' will reach the final state of M and both its heads will be on $. The transitions $U_{\$,\$}|q\rangle=|q_{acc}\rangle$ for all q∈F, takes M' to its accepting state. Thus M' accepts w.

For a string w, which M does not accept, there is no sequence of transitions that takes M to its final state after consumption of w. Thus, no matter what guess the complementarity strings of w make, M' while simulating M based on the complementarity string of w will never reach the situation where both its head is on $ and M' is in a final state of M. Thus the transitions of the form $U_{\$,\$}|q\rangle=|q_{acc}\rangle$ for all q∈F cannot be applied to M'. As a result M' never reaches its accepting state and eventually it will come across two symbols on the two strands for which a transition is not defined as a result M' will go to a rejecting state (As M' is one way and in each state of M' except for the accepting and rejecting states, M' moves both heads to the right, in the worst case both heads of M' will be on '$' and M' is in a state q∈(Q-F) for which no transition is defined, so M' will go to a rejecting state). Thus M' will reject w.

**Example 2:** Consider the deterministic finite automaton M which recognizes the regular language $(a+b)^*a$. This language is not accepted by 1QFA[1]. Hence this language is not reversible which is evident from the transitions of M. Here we will show how to obtain a deterministic Watson-Crick quantum finite automaton M' which recognizes the same regular language.

Let, M=(V, Q, $q_0$, F, δ) where Q={$q_0$, $q_1$}, V={a,b}, $q_0$ is the start state, F={$q_1$} and δ: $q_0(a)\rightarrow q_1$, $q_0(b)\rightarrow q_0$, $q_1(a)\rightarrow q_1$, $q_1(b)\rightarrow q_0$.

The equivalent deterministic Watson-Crick quantum finite automaton M' using the above mentioned procedure is, M'=(Q', V', δ', $q_0$', $Q_{acc}$, $Q_{rej}$, ρ), Q'={$q_0$', $q_0$, $q_1$, $q_{acc}$}, $q_0$' is the start state, $Q_{acc}$={$q_{acc}$}, $Q_{rej}$={}, V'={a, $a_1$, $a_2$, b, $b_1$, $b_2$} and ρ={(a, $a_1$), (a, $a_2$), (b, $b_1$), (b, $b_2$)}. The transitions of M' are as follows:

$U_{\#,\#}|q_0'\rangle=|q_0\rangle$, $U_{a,a_1}|q_0\rangle=|q_1\rangle$, $U_{b,b_1}|q_0\rangle=|q_0\rangle$, $U_{a,a_2}|q_1\rangle=|q_1\rangle$, $U_{b,b_2}|q_1\rangle=|q_0\rangle$, $U_{\$,\$}|q_1\rangle=|q_{acc}\rangle$, D($q_0$')=(1,1), D($q_0$)=(1,1), D($q_1$)=(1,1), D($q_{acc}$)=(0,0).

**Corollary 1:** Watson-Crick quantum finite automata can accept all regular languages.

Proof: From Theorem 2, we know that for every deterministic finite automaton there is a Watson-Crick quantum finite automaton which accepts the same language. For every regular language there is a deterministic finite automaton which accepts that language, thus for every regular languages there is a Watson-Crick quantum finite automaton that accepts it.

**Theorem 3:** The language $L = \{\%w_1*x_1\%w_2*x_2...\%w_n*x_n | n \geq 0, w_i \in \{a,b\}^*, x_i \in \{a,b\}^*, \exists i \exists j : w_i = w_j, x_i \neq x_j\}$ is accepted by a Watson-Crick quantum finite automaton with non-injective complementarity relation.

**Proof:** $M=(Q, V, \delta, q_0, Q_{acc}, Q_{rej}, \rho)$ is a Watson-crick quantum finite automaton that accepts $L=\{\%w_1*x_1\%w_2*x_2...\%w_n*x_n | n \geq 0, w_i \in \{a,b\}^*, x_i \in \{a,b\}^*, \exists i \exists j : w_i = w_j, x_i \neq x_j\}$ where $Q=\{q_0, q_1, q_2, q_3, q_4, q_5\}$, $Q_{acc}=\{q_5\}$, $Q_{rej}=\{q_4\}$, $V=\{a,b,v_{m1},v_{m2},\%,*\}$ $\rho=\{(a,a),(\%,\%),(\%,v_{m1}),(\%,v_{m2}),(b,b),(*,*)\}$.

We define the transitions of M as follows:

$U_{\#,\#}|q_0\rangle=|q_0\rangle$, $U_{\%,\%}|q_0\rangle=|q_0\rangle$, $U_{a,a}|q_0\rangle=|q_0\rangle$, $U_{b,b}|q_0\rangle=|q_0\rangle$, $U_{*,*}|q_0\rangle=|q_0\rangle$, $U_{\%,v_{m1}}|q_0\rangle=|q_1\rangle$, $U_{\%,a}|q_1\rangle=|q_1\rangle$, $U_{\%,b}|q_1\rangle=|q_1\rangle$, $U_{\%,*}|q_1\rangle=|q_1\rangle$, $U_{\%,\%}|q_1\rangle=|q_1\rangle$, $U_{\%,v_{m2}}|q_1\rangle=|q_2\rangle$, $U_{a,a}|q_2\rangle=|q_2\rangle$, $U_{b,b}|q_2\rangle=|q_2\rangle$, $U_{*,*}|q_2\rangle=|q_3\rangle$, $U_{b,b}|q_3\rangle=|q_3\rangle$, $U_{a,a}|q_3\rangle=|q_3\rangle$, $U_{\%,\%}|q_3\rangle=|q_4\rangle$, $U_{\%,\$}|q_3\rangle=|q_4\rangle$, $U_{a,b}|q_3\rangle=|q_5\rangle$, $U_{a,*}|q_3\rangle=|q_5\rangle$, $U_{a,\%}|q_3\rangle=|q_5\rangle$, $U_{a,\$}|q_3\rangle=|q_5\rangle$, $U_{b,a}|q_3\rangle=|q_5\rangle$, $U_{b,*}|q_3\rangle=|q_5\rangle$, $U_{b,\%}|q_3\rangle=|q_5\rangle$, $U_{b,\$}|q_3\rangle=|q_5\rangle$, $U_{*,a}|q_3\rangle=|q_5\rangle$, $U_{*,b}|q_3\rangle=|q_5\rangle$, $U_{*,\%}|q_3\rangle=|q_5\rangle$, $U_{*,\$}|q_3\rangle=|q_5\rangle$, $U_{\%,a}|q_3\rangle=|q_5\rangle$, $U_{\%,b}|q_3\rangle=|q_5\rangle$, $U_{\%,*}|q_3\rangle=|q_5\rangle$.

$D(q_0)=(1,1)$, $D(q_1)=(0,1)$, $D(q_2)=(1,1)$, $D(q_3)=(1,1)$, $D(q_4)=(0,0)$, $D(q_5)=(0,0)$.

The above stated automaton works in the following manner:

The elements $(\%, v_{m1})$ and $(\%, v_{m2})$ of the complementarity relation $\rho$ are used to guess the two substrings of the input string which has its w parts equal and x parts unequal. On finding these guessed substrings the automaton goes to state $q_2$. In state $q_2$, the automaton M checks to see whether the guessed substrings have their w parts equal or not. If the substrings do not have their w parts equal then the automaton halts in a rejecting state as no transitions are defined for such a situation in state $q_2$ and the automaton rejects the input string with probability 1. If the two guessed substring have their w parts equal then the automaton goes to state $q_3$. In state $q_3$, the automaton M checks whether the guessed substrings having their w parts equal have their x parts equal or not. If the x parts are equal then the automaton goes to state $q_4$. The state $q_4$ is a rejecting state, thus, the automaton rejects the input string with probability 1. If the x parts are unequal then the automaton goes to state $q_5$ which is a accepting state; thus the input string is accepted with probability 1 as the guessed substrings have their w parts equal and x parts not equal.

Consider a string s in L. One of the many complementarity strings of s will correctly guess the two substrings which have their w parts equal and x parts unequal and s will be accepted by the automaton M with probability 1.

Now, consider a string s not in L. As s is not in L there are no two substrings of s whose w parts are equal and x parts are unequal. Therefore, no matter the guess made by any complementarity string of s for the location of two substrings of s they will never have their w parts equal and x parts unequal. So M rejects s with probability 1. Thus, from the above stated arguments we conclude that M accepts L.

**Lemma 1:** The language $L = \{\%w_1*x_1\%w_2*x_2...\%w_n*x_n | n \geq 0, w_i \in \{a,b\}^*, x_i \in \{a,b\}^*, \exists i \exists j : w_i = w_j, x_i \neq x_j\}$ is not accepted by any deterministic multi-head finite automaton.
 The proof of Lemma 1 is in Yao et. al.[13]

**Theorem 4:** $L_{1WKQFA} - L_{DFA(k)} \neq \emptyset$, where $L_{1WKQFA}$ is the set of all languages accepted by Watson-Crick quantum finite automata and $L_{DFA(k)}$ is the set of all languages accepted by multi-head deterministic finite automata.

**Proof:** From Theorem 2, we know that there is a Watson-Crick quantum finite automaton that accepts the language $L = \{\%w_1*x_1\%w_2*x_2...\%w_n*x_n | n \geq 0, w_i \in \{a,b\}^*, x_i \in \{a,b\}^*, \exists i \exists j : w_i = w_j, x_i \neq x_j\}$ and from Lemma 1 we know that $L = \{\%w_1*x_1\%w_2*x_2...\%w_n*x_n | n \geq 0, w_i \in \{a,b\}^*, x_i \in \{a,b\}^*, \exists i \exists j : w_i = w_j, x_i \neq x_j\}$ is not accepted by any deterministic multi-head finite automaton which proves the above Theorem.

**Corollary 2:** $L_{1WKQFA} - L_{SDWK} \neq \emptyset$, where $L_{1WKQFA}$ is the set of all languages accepted by Watson-Crick quantum finite automata and $L_{SDWK}$ is the set of all languages accepted by strongly deterministic Watson-Crick automata.

**Proof:** Czeizler et. al.[9] states that the computational power of strongly deterministic Watson-Crick automata and deterministic finite automata with two heads are same and from Theorem 3 we see $L_{1WKQFA} - L_{DFA(k)} \neq \emptyset$ hence the above stated Corollary holds.

**Corollary 3:** The set of languages accepted by one-way reversible multi-head finite automata with two heads is a proper subset of set of languages accepted Watson-Crick quantum finite automata.

**Proof:** From the definition of strongly Watson-Crick quantum finite automata it is evident that if no superposition states are involved in the strongly Watson-Crick quantum finite automaton, then the automaton behaves like a one-way reversible finite automata with two heads, thus for every one-way reversible finite automata with two heads there is a strongly Watson-Crick quantum finite automaton which accepts the same language. Moreover, it has already been stated by Kutrib et.al.[3] that set of languages accepted by multi-head reversible finite automata is a proper subset of set of languages accepted by multi-head deterministic finite automata. Thus there is no multi-head reversible finite automaton which accept the language $L = \{\%w_1*x_1\%w_2*x_2...\%w_n*x_n | n \geq 0, w_i \in \{a,b\}^*, x_i \in \{a,b\}^*, \exists i \exists j : w_i = w_j, x_i \neq x_j\}$ but from Theorem 3, we see that a Watson-Crick quantum automaton can accept the language L, which proves the Corollary.

**Theorem 5:** There is a Watson-Crick quantum finite automaton that accepts the context sensitive language $L = \{ww | w \in \{a,b\}^*\}$.

$M = (Q, V, \delta, q_0, Q_{acc}, Q_{rej}, \rho)$ is a Watson-Crick quantum finite automaton with non-injective complementarity relation $\rho$ that accepts the context sensitive language $L = \{ww | w \in \{a,b\}^*\}$ where $Q = \{q_0, q_1, q_2, q_3, q_4, q_5, q_6, q_7, q_8, q_{rej}, q_{rej1}, q_{rej2}, s_1, s_2\}$, $Q_{acc} = \{s_2\}$, $Q_{rej} = \{s_1, q_{rej}, q_{rej1}, q_{rej2}\}$, $V = \{a, b, m\}$, $\rho = \{(a,a), (a,m), (b,b), (b,m)\}$. We define the transitions involved in M as follows:

$U_{\#,\#}|q_0\rangle = |q_0\rangle$, $U_{\#,a}|q_0\rangle = |q_0\rangle$, $U_{\#,b}|q_0\rangle = |q_0\rangle$, $U_{\#,m}|q_0\rangle = \frac{1}{\sqrt{2}}|q_1\rangle + \frac{1}{\sqrt{2}}|q_2\rangle$, $U_{\#,m}|q_1\rangle = |q_3\rangle$, $U_{a,a}|q_3\rangle = |q_3\rangle$, $U_{b,b}|q_3\rangle = |q_3\rangle$, $U_{a,b}|q_3\rangle = |q_{rej}\rangle$, $U_{b,a}|q_3\rangle = |q_{rej}\rangle$, $U_{x,\$}|q_3\rangle = |q_4\rangle$ $x \in \{a,b\}$, $U_{x,\$}|q_4\rangle = |q_4\rangle$ $x \in \{a,b\}$, $U_{\$,\$}|q_4\rangle = |q_5\rangle$, $U_{\#,m}|q_2\rangle = |q_6\rangle$, $U_{x,y}|q_6\rangle = |q_7\rangle$ $x,y \in \{a,b\}$, $U_{x,y}|q_7\rangle = |q_6\rangle$ $x,y \in \{a,b\}$, $U_{x,\$}|q_6\rangle = |q_{rej2}\rangle$ $x \in \{a,b\}$, $U_{\$,x}|q_6\rangle = |q_{rej2}\rangle$ $x \in \{a,b\}$, $U_{\$,x}|q_7\rangle = |q_{rej1}\rangle$ $x \in \{a,b\}$, $U_{x,\$}|q_7\rangle = |q_{rej1}\rangle$ $x \in \{a,b\}$, $U_{\$,\$}|q_7\rangle = |q_8\rangle$, $U_{\$,\$}|q_5\rangle = \frac{1}{\sqrt{2}} \sum_{l=1}^{2} e^{\frac{2\pi i}{2} \cdot 1 \cdot l} |s_l\rangle$, $U_{\$,\$}|q_8\rangle = \frac{1}{\sqrt{2}} \sum_{l=1}^{2} e^{\frac{2\pi i}{2} \cdot 2 \cdot l} |s_l\rangle$. $D(q_0) = (0,1)$, $D(q_1) = (0,0)$, $D(q_2) = (0,0)$, $D(q_3) = (1,1)$, $D(q_4) = (1,0)$, $D(q_5) = (0,0)$, $D(q_6) = (1,1)$, $D(q_7) = (1,0)$, $D(q_8) = (0,0)$, $D(s_1) = (0,0)$, $D(s_2) = (0,0)$, $D(q_{rej}) = (0,0)$, $D(q_{rej1}) = (0,0)$, $D(q_{rej2}) = (0,0)$.

The above mentioned automaton works in the following manner:

The automaton M works in three phases. In the first phase, the elements (a, m) or (b, m) of the complementarity relation $\rho$ is used to guess the end of first word w in the lower strand. On finding m, the automaton M goes to the second phase. In the second phase the computation branches into 2 paths, indicated by the states $q_1$ and $q_2$ each with amplitude $\frac{1}{\sqrt{2}}$. In each of these two paths, the two tape heads of each individual path move deterministically from the current position to the right end marker '$\$$' independently. The first path checks whether the string in the upper strand after # to the position of m is same as the string in the lower strand after m to $\$$. If some character is not the same then this path ends in rejecting state(i.e. the path verifies whether input is of the form ww ). At the same time, in the second path every time upper head is moved two steps, the lower head is moved one step, to check whether position of 'm' at the end of first word w has been correctly guessed. Only if the position of 'm' is correctly guessed will the two heads of the second path reach '$\$$' at the same time otherwise only one head of the second path goes to '$\$$' and the computation in the second path halts in a rejecting state.

In the third phase, when both the heads of the individual paths arrive at '$\$$' computation in each path again splits according to the quantum Fourier transform yielding either the single accepting state $s_2$ with probability 1 or a rejecting state with probability at least $\frac{1}{2}$.

Now consider a string s in L. As s is in L, s is of the form ww, one of the many complementarity strings of s will guess the position of m at the end of first w correctly, let that string be s'. All the four heads of the two individual paths of automaton M with s in the upper strand and s' in the lower strand will reach '$\$$' at the same time, by the superposition of the machine immediately after performing quantum Fourier transform we get $\frac{1}{2} \sum_{j=1}^{2} \sum_{l=1}^{2} (e^{\frac{2\pi i}{2} \cdot j \cdot l}) |s_l\rangle = |s_2\rangle$. Hence the observable yields the result, accept with probability 1

For a string s not in L, s is not of the form ww, so no matter the position guessed by any complementarity string of w it can never be at the end of first w. As a result, at least one of the 4 heads of two individual paths of M, will not reach $\$$ and the superposition will not result in the accepting state $s_2$. There will be a presence of a rejecting state with probability of atleast $\frac{1}{2}$. Thus M reject w with a probability of at least $\frac{1}{2}$.

## V. CONCLUSION

In this paper, we have introduced a new model of finite automata which combine the features of quantum and DNA computing. We introduced non-determinism in Quantum Computing through the use of non-injective complementarity relation of Watson-Crick automata. We have shown that in spite of Watson-Crick quantum finite automata being reversible in nature they accept all regular languages. We have also explored and compared the computational power Watson-Crick quantum finite automata with other existing deterministic and reversible automata models and utilized the superposition principle to show acceptance of the language $L = \{ww | w \in \{a,b\}^*\}$ by Watson-Crick quantum finite automata. We have also established that the amalgamation of DNA and Quantum model accept languages which are not accepted by any multi-head deterministic finite automata.